\documentclass[twoside,english,final,5p,times,twocolumn]{elsarticle}
\usepackage[T1]{fontenc}
\usepackage[latin9]{inputenc}
\usepackage{amsmath}
\usepackage{amssymb}

\makeatletter
\journal{Journal of Computational Physics}

\makeatother

\usepackage{babel}
\begin{document}

\begin{frontmatter}{}

\title{On the Eigenvalues of the ADER-WENO Galerkin Predictor}

\author{Haran Jackson}

\ead{hj305@cam.ac.uk}

\address{Cavendish Laboratory, JJ Thomson Ave, Cambridge, UK, CB3 0HE}

\end{frontmatter}{}

\section{Background}

ADER-WENO methods have proved extremely useful in obtaining arbitrarily
high-order solutions to problems involving hyperbolic systems of PDEs.
For example, it has been demonstrated that for the same computational
cost as a Runge-Kutta scheme of a certain order, one can obtain an
ADER scheme of one higher order of accuracy (see \citet{Balsara2013}).
Additionally, Runge-Kutta schemes suffer from the presence of Butcher
barriers (see \citet{Butcher2009}), limiting the order of temporal
accuracy that one can comfortably achieve. There are no such limitations
present in ADER-WENO schemes.

The cumbersome analytical derivation of the temporal derivatives of
the solution required by the original ADER formulation (see \citet{Toro2009})
has been replaced by the use of a cell-wise local Galerkin predictor.
The predictor can take either a discontinuous or a continuous form
(see \citet{Dumbser2008} and \citet{Balsara2009}, respectively).
The Galerkin predictor is a high-order polynomial reconstruction of
the data in both space and time, found as the root of a non-linear
system.

It has been conjectured that the eigenvalues of certain matrices appearing
in these non-linear systems are always zero, leading to desirable
system properties for certain classes of PDEs. It is proved here that
this is in deed the case for any number of spatial dimensions and
any desired order of accuracy, for both the discontinuous and continuous
Galerkin variants. This result is independent of the choice of reconstruction
basis polynomials.

The Einstein summation convention is to be assumed throughout this
paper.

\section{The ADER-WENO Method}

Take a non-homogeneous, non-conservative (and for simplicity, one-dimensional)
hyperbolic system of the form:

\begin{equation}
\frac{\partial\boldsymbol{Q}}{\partial t}+\frac{\partial\boldsymbol{F}\left(\boldsymbol{Q}\right)}{\partial x}+B\left(\boldsymbol{Q}\right)\cdot\frac{\partial\boldsymbol{Q}}{\partial x}=\boldsymbol{S}\left(\boldsymbol{Q}\right)\label{eq:NonConservativeSystem}
\end{equation}

where $\boldsymbol{Q}$ is the vector of conserved variables, $\boldsymbol{F}$
is the conservative nonlinear flux, $B$ is the matrix corresponding
to the purely non-conservative component of the system, and $\boldsymbol{S}\left(\boldsymbol{Q}\right)$
is the algebraic source vector.

Take the set of grid points $x_{0}<x_{1}<\ldots<x_{K}$ and define
$\Delta x_{i}=x_{i+1}-x_{i}$. Take the time steps $t_{0}<t_{1}<\ldots$
while defining $\Delta t_{n}=t_{n+1}-t_{n}$. Following the formulations
presented in \citet{Dumbser2009,Dumbser2014,Balsara2009}, the WENO
method and Galerkin method produce at each time step $t_{n}$ a local
polynomial approximation to $\boldsymbol{Q}$ on each space-time cell
$\left[x_{i},x_{i+1}\right]\times\left[t_{n},t_{n+1}\right]$.

Define the scaled space variable:

\begin{equation}
\chi^{i}=\frac{1}{\Delta x_{i}}\left(x-x_{i}\right)
\end{equation}

Take a basis $\left\{ \psi_{0},...,\psi_{N}\right\} $ of $P_{N}$
and inner product $\left\langle \cdot,\cdot\right\rangle $. This
basis can either be nodal ($\psi_{i}\left(\chi_{j}\right)=\delta_{ij}$
where $\left\{ \chi_{0},\ldots,\chi_{N}\right\} $ are a set of nodal
points, such as the Gauss-Legendre abscissae), or modal (such as the
Jacobi polynomials).

The WENO method (as used in \citet{Dumbser2013}) produces an order-$N$
polynomial reconstruction of the data at time $t_{n}$ in cell $\left[x_{i},x_{i+1}\right]$,
using $\left\{ \psi_{0},\ldots,\psi_{N}\right\} $ as a basis. This
is denoted:

\begin{equation}
\boldsymbol{w}\left(x\right)=\boldsymbol{w_{\gamma}}\psi_{\gamma}\left(\chi^{i}\left(x\right)\right)
\end{equation}

This spatial reconstruction at the start of the time step is to be
used as initial data in the problem of finding the Galerkin predictor.

Now define the scaled time variable:

\begin{equation}
\tau^{n}=\frac{1}{\Delta t_{n}}\left(t-t_{n}\right)
\end{equation}

Thus, \eqref{eq:NonConservativeSystem} becomes:

\begin{equation}
\frac{\partial\boldsymbol{Q}}{\partial\tau^{n}}+\frac{\partial\boldsymbol{F^{*}}\left(\boldsymbol{Q}\right)}{\partial\chi^{i}}+B^{*}\left(\boldsymbol{Q}\right)\cdot\frac{\partial\boldsymbol{Q}}{\partial\chi^{i}}=\boldsymbol{S^{*}}\left(\boldsymbol{Q}\right)\label{eq:NonDimensionalNonConservativeSystem}
\end{equation}

where

\begin{equation}
\boldsymbol{F^{*}}=\frac{\Delta t_{n}}{\Delta x}\boldsymbol{F}\qquad B^{*}=\frac{\Delta t_{n}}{\Delta x}B\qquad\boldsymbol{S^{*}}=\Delta t_{n}\boldsymbol{S}
\end{equation}

The non-dimensionalization notation and spacetime cell indexing notation
will be dropped for simplicity in what follows. Now define the set
of spatio-temporal basis functions:

\begin{equation}
\left\{ \theta_{k}\left(\chi,\tau\right)\right\} =\left\{ \psi_{p}\left(\chi\right)\psi_{s}\left(\tau\right):0\leq p,s\leq N\right\} 
\end{equation}

Denoting the Galerkin predictor by $\boldsymbol{q}$, take the following
set of approximations:

\begin{subequations}

\begin{align}
\boldsymbol{Q} & \approx\boldsymbol{q}=\theta_{\beta}\boldsymbol{q_{\beta}}\\
\boldsymbol{F}\left(\boldsymbol{Q}\right) & \approx\theta_{\beta}\boldsymbol{F_{\beta}}\\
B\left(\boldsymbol{Q}\right)\cdot\frac{\partial\boldsymbol{Q}}{\partial\chi} & \approx\theta_{\beta}\boldsymbol{B_{\beta}}\\
\boldsymbol{S}\left(\boldsymbol{Q}\right) & \approx\theta_{\beta}\boldsymbol{S_{\beta}}
\end{align}

\end{subequations}

for some coefficients $\boldsymbol{q_{\beta}},\boldsymbol{F_{\beta}},\boldsymbol{B_{\beta}},\boldsymbol{S_{\beta}}$.

If $\left\{ \psi_{0},...,\psi_{N}\right\} $ is a nodal basis, the
\textit{nodal basis representation} may be used:

\begin{subequations}

\begin{align}
\boldsymbol{F_{\beta}} & =\boldsymbol{F}\left(\boldsymbol{q_{\beta}}\right)\\
\boldsymbol{B_{\beta}} & =B\left(\boldsymbol{q_{\beta}}\right)\cdot\left(\frac{\partial\theta_{\gamma}\left(\chi_{\beta},\tau_{\beta}\right)}{\partial\chi}\boldsymbol{q_{\gamma}}\right)\\
\boldsymbol{S_{\beta}} & =\boldsymbol{S}\left(\boldsymbol{q_{\beta}}\right)
\end{align}

\end{subequations}

where $\left(\chi_{\beta},\tau_{\beta}\right)$ are the coordinates
of the node corresponding to basis function $\theta_{\beta}$.

If a modal basis is used, $\boldsymbol{F_{\beta}},\boldsymbol{B_{\beta}},\boldsymbol{S_{\beta}}$
may be found from the previous values of $\boldsymbol{q_{\beta}}$
in the iterative processes described below.

For functions $f\left(\chi,\tau\right)=f_{\chi}\left(\chi\right)f_{\tau}\left(\tau\right)$
and $g\left(\chi,\tau\right)=g_{\chi}\left(\chi\right)g_{\tau}\left(\tau\right)$,
define the following integral operators:

\begin{subequations}

\begin{align}
\left[f,g\right]^{t} & =f_{\tau}\left(t\right)g_{\tau}\left(t\right)\left\langle f_{\chi},g_{\chi}\right\rangle \\
\left\{ f,g\right\}  & =\left\langle f_{\tau},g_{\tau}\right\rangle \left\langle f_{\chi},g_{\chi}\right\rangle 
\end{align}

\end{subequations}

Multiplying \eqref{eq:NonDimensionalNonConservativeSystem} by test
function $\theta_{\alpha}$, using the polynomial approximations for
$\boldsymbol{Q},\boldsymbol{F},\boldsymbol{B},\boldsymbol{S}$, and
integrating over space and time gives:

\begin{align}
\left\{ \theta_{\alpha},\frac{\partial\theta_{\beta}}{\partial\tau}\right\} \boldsymbol{q_{\beta}} & =-\left\{ \theta_{\alpha},\frac{\partial\theta_{\beta}}{\partial\chi}\right\} \boldsymbol{F_{\beta}}+\left\{ \theta_{\alpha},\theta_{\beta}\right\} \left(\boldsymbol{S_{\beta}}-\boldsymbol{B_{\beta}}\right)\label{eq:InitialSystem}
\end{align}

\subsection{The Discontinuous Galerkin Method}

This method of computing the Galerkin predictor allows solutions to
be discontinuous at temporal cell boundaries, and is also suitable
for stiff source terms.

Integrating \eqref{eq:InitialSystem} by parts in time gives:

\begin{align}
\left(\left[\theta_{\alpha},\theta_{\beta}\right]^{1}-\left\{ \frac{\partial\theta_{\alpha}}{\partial\tau},\theta_{\beta}\right\} \right)\boldsymbol{q_{\beta}} & =\left[\theta_{\alpha},\boldsymbol{w}\right]^{0}-\left\{ \theta_{\alpha},\frac{\partial\theta_{\beta}}{\partial\chi}\right\} \boldsymbol{F_{\beta}}\\
 & +\left\{ \theta_{\alpha},\theta_{\beta}\right\} \left(\boldsymbol{S_{\beta}}-\boldsymbol{B_{\beta}}\right)\nonumber 
\end{align}

where $\boldsymbol{w}$ is the reconstruction obtained at the start
of the time step with the WENO method. Define the following:

\begin{subequations}

\begin{align}
U_{\alpha\beta} & =\left[\theta_{\alpha},\theta_{\beta}\right]^{1}-\left\{ \frac{\partial\theta_{\alpha}}{\partial\tau},\theta_{\beta}\right\} \\
V_{\alpha\beta} & =\left\{ \theta_{\alpha},\frac{\partial\theta_{\beta}}{\partial\chi}\right\} \\
\boldsymbol{W_{\alpha}} & =\left[\theta_{\alpha},\psi_{\gamma}\right]^{0}\boldsymbol{w_{\gamma}}\\
Z_{\alpha\beta} & =\left\{ \theta_{\alpha},\theta_{\beta}\right\} 
\end{align}

\end{subequations}

Thus:

\begin{equation}
U_{\alpha\beta}\boldsymbol{q_{\beta}}=\boldsymbol{W_{\alpha}}-V_{\alpha\beta}\boldsymbol{F_{\beta}}+Z_{\alpha\beta}\left(\boldsymbol{S_{\beta}}-\boldsymbol{B_{\beta}}\right)
\end{equation}

This nonlinear system in $\boldsymbol{q_{\beta}}$ is solved by a
Newton method. The source terms must be solved implicitly if they
are stiff. Note that $\boldsymbol{W}$ has no dependence on $\boldsymbol{q}$.

\subsection{The Continuous Galerkin Method}

This method of computing the Galerkin predictor is not suitable for
stiff source terms, but it provides substantial savings on computational
cost and ensures continuity across temporal cell boundaries.

$\left\{ \psi_{0},...,\psi_{N}\right\} $ must be chosen in such a
way that the first $N+1$ elements of $\left\{ \theta_{\beta}\right\} $
have only a spatial dependence. The first $N+1$ elements of $\boldsymbol{q}$
are then fixed by demanding continuity at $\tau=0$:

\begin{equation}
\boldsymbol{q}\left(\chi,0\right)=\boldsymbol{w}\left(\chi\right)
\end{equation}

where $\boldsymbol{w}$ is spatial the reconstruction obtained at
the start of the time step with the WENO method.

For a given vector $\boldsymbol{v}\in\mathbb{R}^{\left(N+1\right)^{2}}$
and matrix $X\in M_{\left(N+1\right)^{2},\left(N+1\right)^{2}}\left(\mathbb{R}\right)$,
let $\boldsymbol{v}=\left(\boldsymbol{v^{0}},\boldsymbol{v^{1}}\right)$
and $X=\left(\begin{array}{cc}
X^{00} & X^{01}\\
X^{10} & X^{11}
\end{array}\right)$ where $\boldsymbol{v^{0}},X^{00}$ are the components relating solely
to the first $N+1$ components of $\boldsymbol{v}$. We only need
to find the latter components of $\boldsymbol{q}$, and thus, from
\eqref{eq:InitialSystem}, we have:

\begin{align}
\left\{ \theta_{\alpha},\frac{\partial\theta_{\beta}}{\partial\tau}\right\} ^{11}\boldsymbol{q_{\beta}^{1}} & =\left\{ \theta_{\alpha},\theta_{\beta}\right\} ^{11}\left(\boldsymbol{S_{\beta}^{1}}-\boldsymbol{B_{\beta}^{1}}\right)-\left\{ \theta_{\alpha},\frac{\partial\theta_{\beta}}{\partial\chi}\right\} ^{11}\boldsymbol{F_{\beta}^{1}}\\
 & +\left\{ \theta_{\alpha},\theta_{\beta}\right\} ^{10}\left(\boldsymbol{S_{\beta}^{0}}-\boldsymbol{B_{\beta}^{0}}\right)-\left\{ \theta_{\alpha},\frac{\partial\theta_{\beta}}{\partial\chi}\right\} ^{10}\boldsymbol{F_{\beta}^{0}}\nonumber 
\end{align}
Define the following:

\begin{subequations}

\begin{align}
U_{\alpha\beta} & =\left\{ \theta_{\alpha},\frac{\partial\theta_{\beta}}{\partial\tau}\right\} ^{11}\\
V_{\alpha\beta} & =\left\{ \theta_{\alpha},\frac{\partial\theta_{\beta}}{\partial\chi}\right\} ^{11}\\
\boldsymbol{W_{\alpha}} & =\left\{ \theta_{\alpha},\theta_{\beta}\right\} ^{10}\left(\boldsymbol{S_{\beta}}-\boldsymbol{B_{\beta}}\right)^{0}-\left\{ \theta_{\alpha},\frac{\partial\theta_{\beta}}{\partial\chi}\right\} ^{10}\boldsymbol{F_{\beta}^{0}}\\
Z_{\alpha\beta} & =\left\{ \theta_{\alpha},\theta_{\beta}\right\} ^{11}
\end{align}

\end{subequations}

Thus:

\begin{equation}
U_{\alpha\beta}\boldsymbol{q_{\beta}^{1}}=\boldsymbol{W_{\alpha}}-V_{\alpha\beta}\boldsymbol{F_{\beta}^{1}}+Z_{\alpha\beta}\left(\boldsymbol{S_{\beta}^{1}}-\boldsymbol{B_{\beta}^{1}}\right)
\end{equation}

Note that, as with the discontinuous Galerkin method, $\boldsymbol{W}$
has no dependence on the degrees of freedom in $\boldsymbol{q}$.

\section{Conjecture}

Extending the Galerkin method described in the previous section to
three dimensions, the following system must be solved for $\boldsymbol{q}$:

\begin{align}
U_{\alpha\beta}\boldsymbol{q_{\beta}} & =\boldsymbol{W_{\alpha}}-V_{\alpha\beta}^{1}\boldsymbol{F_{\beta}}-V_{\alpha\beta}^{2}\boldsymbol{G_{\beta}}-V_{\alpha\beta}^{3}\boldsymbol{H_{\beta}}\label{eq:DGIteration}\\
 & +Z_{\alpha\beta}\left(\boldsymbol{S_{\beta}}-\boldsymbol{B_{\beta}}\right)\nonumber 
\end{align}

where now we have the 3 scaled spatial variables $\chi_{1},\chi_{2},\chi_{3}$
and $\boldsymbol{G,H}$ are the flux components in the second and
third spatial directions, respectively. In the case of the continuous
Galerkin method, it is assumed that \eqref{eq:DGIteration} is to
be solved for only the non-fixed degrees of freedom in $\boldsymbol{q}$.
The matrices $V_{\alpha\beta}^{i}$ are defined thus:

\begin{equation}
V_{\alpha\beta}^{i}=\left\langle \theta_{\alpha},\frac{\partial\theta_{\beta}}{\partial\chi_{i}}\right\rangle 
\end{equation}

For the discontinuous Galerkin method, $\boldsymbol{W_{\alpha}}$
now takes the form:

\begin{equation}
\boldsymbol{W_{\alpha}}=\left[\theta_{\alpha},\Psi_{\gamma}\right]^{0}\boldsymbol{w_{\gamma}}
\end{equation}

where $\Psi_{\gamma}\left(\chi_{1},\chi_{2},\chi_{3}\right)$ is an
element of the following set, enumerated by $\gamma$:

\begin{equation}
\left\{ \psi_{i}\left(\chi_{1}\right)\psi_{j}\left(\chi_{2}\right)\psi_{c}\left(\chi_{3}\right):0\leq i,j,k\leq N\right\} 
\end{equation}

For the continuous Galerkin method, $\boldsymbol{W_{\alpha}}$ now
takes the form:

\begin{align}
\boldsymbol{W_{\alpha}} & =\left\{ \theta_{\alpha},\theta_{\beta}\right\} ^{10}\left(\boldsymbol{S_{\beta}}-\boldsymbol{B_{\beta}}\right)^{0}\\
 & -\left\{ \theta_{\alpha},\frac{\partial\theta_{\beta}}{\partial\chi^{1}}\right\} ^{10}\boldsymbol{F_{\beta}^{0}}-\left\{ \theta_{\alpha},\frac{\partial\theta_{\beta}}{\partial\chi^{2}}\right\} ^{10}\boldsymbol{G_{\beta}^{0}}-\left\{ \theta_{\alpha},\frac{\partial\theta_{\beta}}{\partial\chi^{2}}\right\} ^{10}\boldsymbol{H_{\beta}^{0}}\nonumber 
\end{align}

\citet{Dumbser2008} remark that for the continuous Galerkin case,
the eigenvalues of $U^{-1}V^{i}$ are all 0 for $0\leq N\leq5$, for
$i=1,2,3$. \citet{Dumbser2009a} state the same result for the discontinuous
Galerkin case. This implies that in the conservative, homogeneous
case ($\boldsymbol{B}=\boldsymbol{S}=\boldsymbol{0}$), owing to the
Banach Fixed Point Theorem, existence and uniqueness of a solution
are established, and convergence to this solution is guaranteed. As
noted in \citet{Dumbser2009a}, in the linear case it is implied that
the iterative procedure converges after at most $N+1$ iterations.

In \citet{Dumbser2008} it is conjectured that the result concerning
the eigenvalues of $U^{-1}V^{i}$ holds for any $N$, and any number
of spatial dimensions. A solution to this conjecture is provided here.
For the theory in linear algebra required for this section, please
consult a standard textbook on the subject, such as \citet{Nering1970}.

\subsection{The Discontinuous Galerkin Case}

First, given the basis polynomials $\left\{ \psi_{0},\ldots,\psi_{N}\right\} $,
define the following matrices:

\begin{subequations}

\begin{align}
\aleph_{ij} & =\left\langle \psi_{i},\psi_{j}\right\rangle \\
\beth_{ij} & =\left\langle \psi_{i},\psi_{j}^{'}\right\rangle 
\end{align}

\end{subequations}

Note that $\aleph$ is the Gram matrix, which by linear independence
of $\left\{ \psi_{0},...,\psi_{N}\right\} $ is invertible. Note also
that if $p\in P_{N}$ then $p=a_{j}\psi_{j}$ for some unique coefficient
vector $\boldsymbol{a}$. Thus, taking inner products with $\psi_{i}$,
we have $\left\langle \psi_{i},\psi_{j}\right\rangle a_{j}=\left\langle \psi_{i},p\right\rangle $
for $i=0,...,N$. This produces the following result:

\begin{equation}
p=a_{j}\psi_{j}\Leftrightarrow\boldsymbol{a}=\aleph^{-1}\boldsymbol{x},\;x_{i}=\left\langle \psi_{i},p\right\rangle \label{eq:gramCoeffs}
\end{equation}

Without loss of generality, take the ordering:

\begin{equation}
\alpha=\alpha_{t}\left(N+1\right)^{3}+\alpha_{x}\left(N+1\right)^{2}+\alpha_{y}\left(N+1\right)+\alpha_{z}
\end{equation}

where $0\leq\alpha_{t},\alpha_{x},\alpha_{y},\alpha_{z}\leq N$. Using
the same ordering for $\beta$, we have:

\begin{subequations}

\begin{align}
U_{\alpha\beta} & =\left(\psi_{\alpha_{t}}\left(1\right)\psi_{\beta_{t}}\left(1\right)-\beth_{\beta_{t}\alpha_{t}}\right)\cdot\aleph_{\alpha_{x}\beta_{x}}\cdot\aleph_{\alpha_{y}\beta_{y}}\cdot\aleph_{\alpha_{z}\beta_{z}}\\
V_{\alpha\beta}^{1} & =\aleph_{\alpha_{t}\beta_{t}}\cdot\beth_{\alpha_{x}\beta_{x}}\cdot\aleph_{\alpha_{y}\beta_{y}}\cdot\aleph_{\alpha_{z}\beta_{z}}
\end{align}

\end{subequations}

Thus:

\begin{subequations}

\begin{align}
U & =C\otimes\aleph\otimes\aleph\otimes\aleph\\
V^{1} & =\aleph\otimes\beth\otimes\aleph\otimes\aleph
\end{align}

\end{subequations}

where $C_{ij}=\psi_{i}\left(1\right)\psi_{j}\left(1\right)-\beth_{ji}$.
Thus:

\begin{equation}
U^{-1}V^{1}=\left(C^{-1}\aleph\right)\otimes\left(\aleph^{-1}\beth\right)\otimes I\otimes I
\end{equation}

Therefore:

\begin{equation}
\left(U^{-1}V^{1}\right)^{k}=\left(C^{-1}\aleph\right)^{k}\otimes\left(\aleph^{-1}\beth\right)^{k}\otimes I\otimes I
\end{equation}

A matrix $X$ is nilpotent ($X^{k}=0$ for some $k\in\mathbb{N}$)
if and only if all its eigenvalues are 0. The conjecture will be proved
if it is shown that $\left(\aleph^{-1}\beth\right)^{k}=0$ for some
$k\in\mathbb{N}$, as this would imply that $\left(U^{-1}V^{1}\right)^{k}=0$,
and thus all eigenvalues of $U^{-1}V^{1}$ are 0.

Taking $\boldsymbol{a}\in\mathbb{R}^{N+1}$, define:

\begin{equation}
p=a_{0}\psi_{0}+\ldots+a_{N}\psi_{N}\in P_{N}
\end{equation}

Note that:

\begin{equation}
\left(\beth\boldsymbol{a}\right)_{i}=\left\langle \psi_{i},\psi_{0}^{'}\right\rangle a_{0}+\ldots+\left\langle \psi_{i},\psi_{N}^{'}\right\rangle a_{N}=\left\langle \psi_{i},p^{'}\right\rangle 
\end{equation}

Thus, by \eqref{eq:gramCoeffs}:

\begin{equation}
\left(\aleph^{-1}\beth\boldsymbol{a}\right)_{i}\psi_{i}=p'
\end{equation}

By induction:

\begin{equation}
\left(\left(\aleph^{-1}\beth\right)^{k}\boldsymbol{a}\right)_{i}\psi_{i}=p^{\left(k\right)}
\end{equation}

for any $k\in\mathbb{N}$. As $p\in P_{N}$, $\left(\aleph^{-1}\beth\right)^{N+1}\boldsymbol{a}=\boldsymbol{0}$.
As $\boldsymbol{a}$ was chosen arbitrarily, $\left(\aleph^{-1}\beth\right)^{N+1}=0$.
Thus, the conjecture is solved.

This proof is easily adapted to show that $U^{-1}V^{2}$ and $U^{-1}V^{3}$
are nilpotent, and clearly extends to any number of spatial dimensions.
No specific choice has been made for $N\in\mathbb{N}$ and thus the
result holds in general.

\subsection{The Continuous Galerkin Case}

In addition to $\aleph,\beth$, we now define $\aleph',\beth'$ where
each new matrix is equal to the original, with its first row and column
removed (the row and column corresponding to the constant-term polynomial
$\psi_{0}$). Take the following ordering:

\begin{equation}
\alpha=\alpha_{t}\left(N+1\right)^{3}+\alpha_{x}\left(N+1\right)^{2}+\alpha_{y}\left(N+1\right)+\alpha_{z}
\end{equation}

where now $0\leq\alpha_{x},\alpha_{y},\alpha_{z}\leq N$ and $0\leq\alpha_{t}\leq N-1$.
Using the same ordering for $\beta$, we now have:

\begin{subequations}

\begin{align}
U_{\alpha\beta} & =\beth'_{\alpha_{t}\beta_{t}}\cdot\aleph_{\alpha_{x}\beta_{x}}\cdot\aleph_{\alpha_{y}\beta_{y}}\cdot\aleph_{\alpha_{z}\beta_{z}}\\
V_{\alpha\beta}^{1} & =\aleph'_{\alpha_{t}\beta_{t}}\cdot\beth_{\alpha_{x}\beta_{x}}\cdot\aleph_{\alpha_{y}\beta_{y}}\cdot\aleph_{\alpha_{z}\beta_{z}}
\end{align}

\end{subequations}

The proof for the continuous case follows in the same manner as the
proof for the discontinuous case, with:

\begin{equation}
U^{-1}V^{1}=\left(\left(\beth'\right)^{-1}\aleph'\right)\otimes\left(\aleph^{-1}\beth\right)\otimes I\otimes I
\end{equation}

\section{References}

\bibliographystyle{elsarticle-harv}
\addcontentsline{toc}{section}{\refname}\bibliography{ref/refs}

\section{Acknowledgments}

I acknowledge financial support from the EPSRC Centre for Doctoral
Training in Computational Methods for Materials Science under grant
EP/L015552/1.
\end{document}